\documentclass[aps,prl,twocolumn,groupedaddress,showpacs]{revtex4}
\usepackage{graphics}
\usepackage{color}

\usepackage[]{graphicx}
\DeclareGraphicsExtensions{.pdf}

\newcommand {\e}[1]{\mathrm{~#1}}       

\begin{document}

\title{Stochastic proofreading mechanism alleviates crosstalk in transcriptional regulation}

\author{Sarah A. Cepeda-Humerez, Georg Rieckh, Ga\v{s}per Tka\v{c}ik}
\affiliation{$^a$Institute of Science and Technology Austria\\
Am Campus 1, A-3400 Klosterneuburg, Austria}

\date{\today}

\begin{abstract}
Gene expression is controlled primarily by interactions between transcription factor proteins (TFs) and the regulatory DNA sequence, a process that can be captured well by thermodynamic models of regulation. These models, however, neglect regulatory crosstalk: the possibility that non-cognate TFs could initiate transcription, with potentially disastrous effects for the cell. Here we estimate the importance of crosstalk, suggest that its avoidance strongly constrains equilibrium models of TF binding, and propose an alternative non-equilibrium scheme that implements kinetic proofreading to suppress erroneous initiation. This proposal is consistent with the observed covalent modifications of the transcriptional apparatus and would predict increased noise in gene expression as a tradeoff for improved specificity. Using information theory, we quantify this tradeoff to find when optimal proofreading architectures are favored over their equilibrium counterparts.
\end{abstract}

\pacs{}
\maketitle
In prokaryotes, transcription factors recognize and bind specific DNA sequences $L=10-20$ basepairs (bp) in length, usually located in promoter regions upstream of the regulated genes \cite{ptashne+gann_02}. 
Regulation by a single TF, or a small number of TFs interacting cooperatively, is sufficient to quantitatively account for the experimental measurements of gene expression \cite{kuhlman+al_07}, as well as to explain how any gene can be individually ``addressed'' and regulated only by its cognate TFs \cite{wunderlich+mirny_09}, without much danger of regulatory crosstalk. 
In eukaryotes, however, TFs seem to be much less specific ($L=5-10$ bp; but the total genome size is larger than in prokaryotes by $\sim 10^3$) \cite{wunderlich+mirny_09,jasper}, binding promiscuously to many genomic locations \cite{li+al_08}, including to their non-cognate binding sites \cite{rockel+al_13}. What are the implications of this reduced specificity for the precision of gene regulation?

Thermodynamic models of regulation postulate that the rate of target gene expression is given by the equilibrium occupancy of various TFs on the regulatory sequence \cite{shea+ackers_84,bintu+al_05}, and the success of this framework in prokaryotes \cite{kinney+al_10} has prompted its application to eukaryotic, in particular, metazoan, enhancers \cite{janssens+al_06,he+al_10,fakhouri+al_10}.  
To illustrate the crosstalk problem in this setting, consider the ratio $\sigma$ of the dissociation constants to a nonspecific and a specific site for an eukaryotic TF; typically, $\sigma \sim 10^3$ (corresponding to  a difference in binding energy of $\sim 7\,k_BT$) \cite{maerkl+quake_07,rockel+al_13}. Because there are $\nu\sim10^2-10^3$ of different TF species in a cell, TFs nonspecific to a given site will greatly outnumber the specific ones. For an isolated binding site, this would imply roughly equal occupancy by cognate and noncognate TFs, suggesting that crosstalk could be acute. For multiple sites, cooperative binding is known for its role in facilitating sharp and strong gene activation even with cognate TFs of intermediate specificity---but could the same mechanism also alleviate crosstalk? First, note that there exist well-studied TFs which do not bind cooperatively (e.g. \cite{giorgetti+al_10}). Second, while many proposed regulation schemes give rise to cooperativity (e.g., nucleosome-mediated cooperativity \cite{mirny_10}, or synergistic activation \cite{todeschini+al_14}) they will not suppress crosstalk; for the latter, cooperativity needs to be strong and specific, stabilizing only the binding of \emph{cognate} TFs. Third, even when cooperative interactions are specific, crosstalk can pose a serious constraint. Regulating a gene implies varying the cognate TF concentration throughout its dynamic range, and when this concentration is low and the target gene should be uninduced, cooperativity  cannot prevent the erroneous induction by noncognate TFs. For that, the cell could either keep the genes inactive by binding of specific repressors, or by making the whole gene unavailable for transcription. The first strategy seems widely used in bacteria but less so in eukaryotes; the second strategy (``gene silencing'') is widespread in eukaryotes, but only happens at a slow timescale and involves a complex series of nonequilibrium steps. 

Here we propose a plausible and fast molecular mechanism which alleviates the effects of crosstalk; a detailed account of when crosstalk poses a severe constraint for gene regulation will be  presented elsewhere. The proposed mechanism is consistent with the known tight control over which genes are expressed in different conditions or tissues (e.g., during development \cite{mcginnis+krumlauf_92}) on the one hand, and on the other, explains  the high levels of measured noise in transcription initiation of active genes \cite{raj+al_06,little+al_13}.

\begin{figure}
\centering
\includegraphics[width = 3.3in]{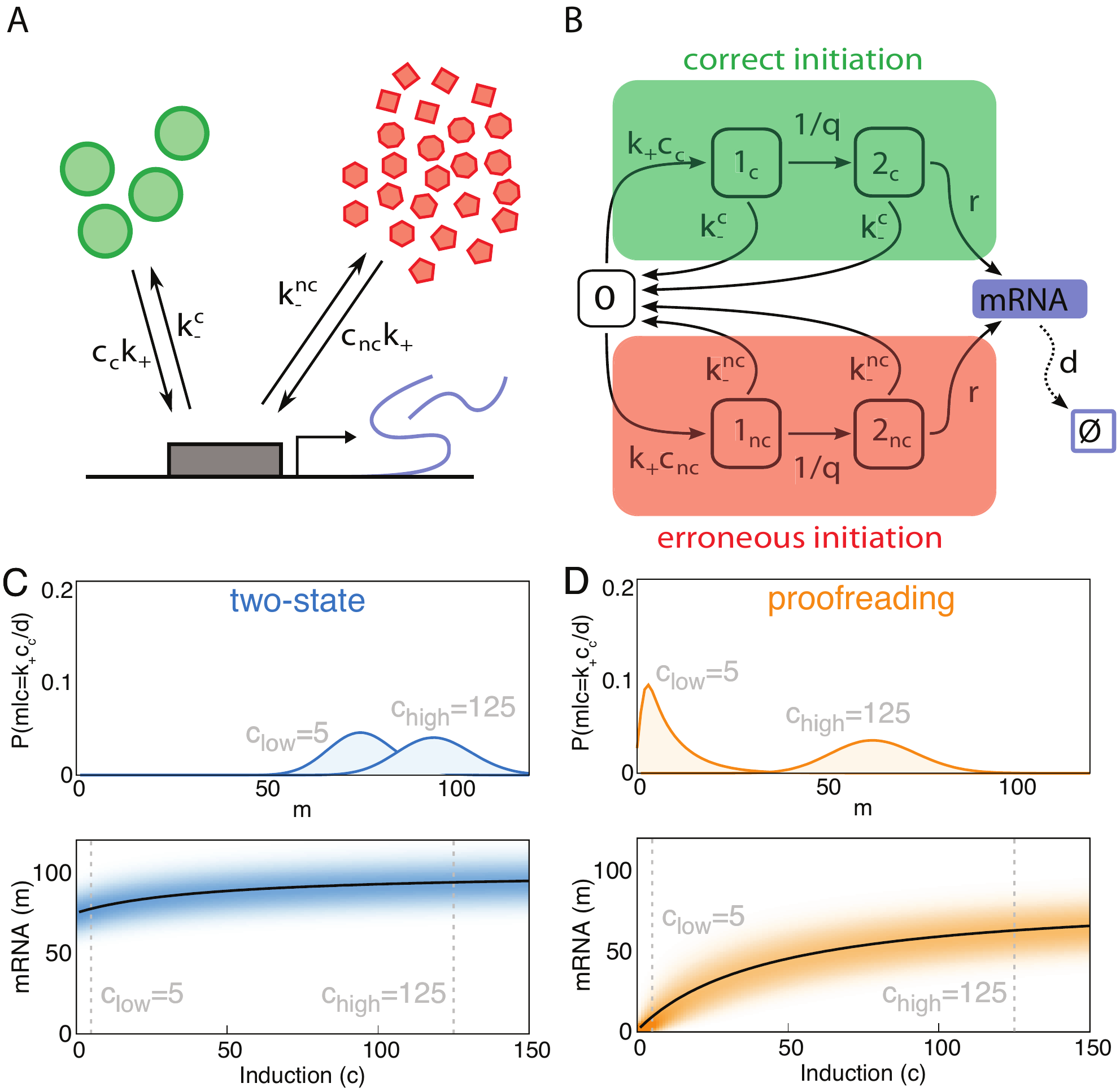}
\caption{{\bf A)} A schematic of cognate (green circles) and $\nu$ kinds of noncognate (various red shapes) TFs binding to a gene regulatory element on the DNA (gray box), to control the mRNA expression level.
{\bf B)} Transition state diagram for the proofreading gene regulation. The regulatory element can cycle between an empty state ($\mathtt{0}$), state occupied by either cognate ($\mathtt{1}_{\rm c}$) or noncognate ($\mathtt{1}_{\rm nc}$) TF; to initiate gene expression, a further non-equilibrium transition into ``$\mathtt{2}$'' states (with rate $1/q$) is required, driven by, e.g., hydrolysis of ATP. mRNA is expressed at rate $r$ and degraded with rate $d$, the slowest process that sets our unit for time.
In this figure we use $r/d=100,k_-^{\rm nc}/d=2500,\sigma=500,\nu=50,\Lambda=\nu/\sigma=0.1$; dimensionless concentration is $c=k_+ c_{\rm c}/d$. {\bf C,D)} Steady-state mRNA distributions for low and high concentrations of the cognate TF, $c$. As $qd\rightarrow 0$ (C), the proofreading model reduces to the two-state model of gene expression \cite{rieckh}; here, noncognate TFs initiate transcription at a high rate even when $c$ is low, causing overlapping output distributions (blue; top) and small dynamic range (black line = $\langle m(c)\rangle$, blue shade = $\sigma_m(c)$; bottom). Proofreading (D) suppresses erroneous initiation, leading to separable output distributions (orange; top) and higher dynamic range (bottom).}
\label{f1}
\end{figure}

The simplest proofreading architecture for transcriptional gene activation that can cope with erroneous binding is presented in Fig~\ref{f1}A,B, motivated by a scheme first proposed by Hopfield \cite{hopfield_74}. Specificity is only conveyed by differential rates of TF unbinding (``off-rates'' $k_-^{\rm c}, k_-^{\rm nc}$, with $\sigma=k_-^{\rm nc}/k_-^{\rm c}$). There are $\nu$ noncognate TF species whose typical concentration we take to be $c_{\rm nc}=\frac{1}{2}\nu C$, and $C$ is the maximal concentration for the cognate TFs $c_{\rm c}$, $c_{\rm c}\in[0,C]$. The ratio $\Lambda = \nu/\sigma$ determines the severity of crosstalk, which is weak for $\Lambda\ll 1$ and strong for $\Lambda\gg 1$. The response of the promoter to the dimensionless input concentration $c$ ($=k_+c_{\rm c}/d$, see Fig~\ref{f1}B) of cognate TFs is captured by the steady state distribution of mRNA, $P(m|c)$; the spread of this distribution is due to the stochasticity in gene expression, which includes random switching between promoter states and the birth-death process of mRNA expression \cite{noise}. If the reaction rates are known, $P(m|c)$ is computable from the chemical Master equation corresponding to the transition diagram in Fig~\ref{f1}B; using finite-state truncation, this becomes a linear problem that is numerically tractable.

Figures~\ref{f1}C and D each compare the steady state distributions of mRNA at low and high concentration of cognate TF, $c$. The behavior crucially depends on the out-of-equilibrium rate $qd$. When $qd\rightarrow 0$, the scheme of Fig~\ref{f1}B becomes a normal two-state promoter as the states $\mathtt{1}_{\rm c}$ and $\mathtt{2}_{\rm c}$ (likewise $\mathtt{1}_{\rm nc}$ and $\mathtt{2}_{\rm nc}$) fuse into a single state. In this limit, the effect of crosstalk is highly detrimental already at $\Lambda=0.1$ used in this example: at low $c$, the promoter repeatedly cycles through erroneous initiation and the gene is highly expressed both at low $c$ as well as at high $c$ (where most of the expression is indeed due to correct initiation); as a result, the distributions $P(m|c)$ show substantial overlap  in the two input conditions shown in Fig~\ref{f1}C. In contrast, for a non-trivial choice of $q$ ($k_-^{\rm c} \ll 1/q \simeq k_-^{\rm nc}$), the model can exhibit proofreading. Even at low cognate concentration $c$, the slow irreversible transition ensures that noncognate TFs unbind from the promoter and that erroneous initiation is consequently rare, which is manifested as a sharp peak of $P(m|c_{\rm low})$ at small $m$ in Fig~\ref{f1}D. The proofreading architecture generates a larger output dynamic range and consequently makes the responses distinguishable.

\begin{figure}
\centering
\includegraphics[width = 3.3in]{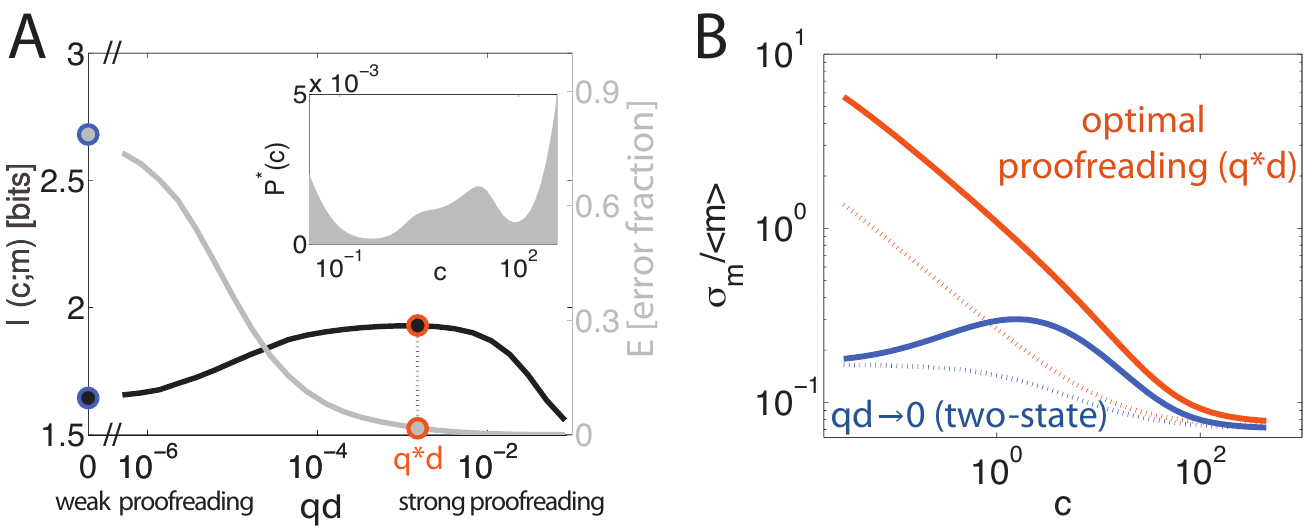}
\caption{{\bf A)} Maximal information transmission (left axis, black) and the error fraction (right axis, gray) as a function of the inverse irreversible reaction rate, $qd$. Increasing $qd$ suppresses the error fraction, but only at the cost of increasing the gene expression noise, leading to a tradeoff and an information-maximizing value of $q^*d$ (orange). This maximum is reached robustly with  input distributions that are close to optimal (inset). {\bf B)} Noise in gene expression, $\sigma_m/\langle m\rangle$, computed from the moments of $P(m|c)$, as a function of the dimensionless input concentration $c$, for the optimal proofreading (orange) and the two-state (blue) architectures. Dotted lines show the Poisson limit, $\sigma_m^2=\langle m\rangle$, for comparison. In both cases, the average number of mRNA expressed if fixed to $\bar{m}=100$.}
\label{f2}
\end{figure}

What are the costs to the cell of the proposed proofreading mechanism? First, the mechanism requires an energy source, e.g., ATP, to break detailed balance. Whether such a metabolic cost is a burden to the cell is unclear: a few molecules of ATP paid per initiation should be negligible compared to the processive cost of transcription and translation. Second, however, is an indirect cost in terms of gene expression noise. While proofreading decreases erroneous induction, it takes longer to traverse the state transition diagram from empty state $\mathtt{0}$ to expressing state $\mathtt{2}$, and since the promoter can perform aborted erroneous initiation cycles, the fluctuations in the time-to-induction will also increase \cite{bel+al_10}. This will result in additional variance in the mRNA copy number at steady state compared to the two-state ($qd\rightarrow 0$) scheme. While  the speed/specificity tradeoff in protein synthesis has been examined before using deterministic chemical kinetics \cite{savir+tlusty}, this stochastic formulation of proofreading has, to our knowledge, remained unexplored. Proofreading in gene regulation is thus expected to increase the output dynamic range, which is favorable for signaling, but also to increase the noise, which is detrimental. 

How can we formalize the tradeoff between noise and dynamic range for gene regulatory schemes and find when proofreading is beneficial? In existing analyses of proofreading the erroneous incorporation of the substrate leads to an error product that is \emph{different} from the correct one \cite{hopfield_74,savir+tlusty}; in contrast, here the gene always expresses the \emph{same} mRNA. What is important for signal transduction, however, is how well this expression correlates with the input signal, $c$. To quantify the regulatory power of the proofreading architecture, we computed the mutual information, $I(c;m)$ \cite{shannon}, between the signal $c$ and the mRNA expression level $m$, following previous applications of information theory to gene regulation \cite{tkacik+walczak_review,rieckh}. The information depends not only on $P(m|c)$, which we compute from the Master equation, but also on the \emph{a priori} unknown distribution of input concentrations, $P(c)$; we therefore determined the input distribution $P^*(c)$ that maximizes information transmission, subject to a constraint on the average number of expressed mRNA, $\bar{m}=\int dc P(c)\sum_m m P(m|c)$. This constraint on average number of mRNA was imposed to compare different regulatory architectures; otherwise, higher average expression could yield higher information transmission for trivial reasons. Such constrained information (capacity) maximization is a well-known problem in information theory that can be solved using the Blahut-Arimoto algorithm \cite{ba}. 

Figure~\ref{f2}A shows how the information transmission $I(m;c)$ through the promoter depends on the (inverse)  reaction rate $qd$.  We start by looking at the classic measure of proofreading performance, the ``error fraction,'' i.e.,  the ratio of the mRNA expressed from state $\mathtt{2}_{\rm nc}$ due to noncognate TFs, vs mRNA expressed from state $\mathtt{2}_{\rm c}$  due to cognate TFs. As $qd$ is increased, the error fraction drops, with no clear optimum. In contrast, there exists an optimal $q^*d$ at which the information is maximized---this is the point where proofreading is most effective, optimally trading off erroneous induction (here, suppressed by a factor of $\sim 30$ relative to no proofreading), noise in gene expression, and dynamic range at the output. In Fig~\ref{f2}B we plot the noise in gene expression, as a function of the input concentration $c$ for the optimal proofreading architecture and the non-proofreading limit. In both cases the noise has super-Poisson components due to the switching between promoter states, but this excess is substantially higher in the proofreading architecture, as expected.

While attractive, these results still depend on the particular rates chosen for the model in Fig~\ref{f1}B. Surprisingly, if we choose to compare the \emph{optimal} proofreading scenario with the \emph{optimal} non-proofreading one, the problem simplifies further. Given that the input TF concentration $c$  varies over some limited dynamic range, $c\in[0,C_{\rm max}=k_+C/d]$, there should exist also an optimal setting for $k_-^{\rm c}$: set too high, the cognate TFs will be extremely unlikely to occupy the promoter for any significant fraction of the time and induce the gene; set too low, the switching contribution to noise in gene expression will blow up. With $k_-^{\rm c}$ and $q$ in the ``correct initiation'' pathway  of Fig~\ref{f1}B set by optimization, the remaining rates in the ``erroneous initiation'' pathway are fixed by the choice of crosstalk severity $\Lambda$. The remaining parameters regulating mRNA expression---the average mRNA count $\bar{m}$ and the rate $r$---do not change the results qualitatively. The mRNA expression rate $r$ simply sets the maximal number of mRNA molecules at full expression in steady state ($r/d$); this influences the Poisson noise at the output, but does so equally for any regulatory architecture, proofreading or not. As long as $r$ is large enough so that the average mRNA constraint $\bar{m}$ is achievable, the precise choice of these values is not crucial (we use $r/d=200$, $\bar{m}=100$, plausible for eukaryotic expression). In sum, we can compare how well the optimal proofreading architecture does compared to optimal non-proofreading architecture in terms of information transmission, as  a function of two key parameters: the crosstalk severity, $\Lambda$, and the input dynamic range, $C_{\rm max}$.

\begin{figure}
\centering
\includegraphics[width = 3.3in]{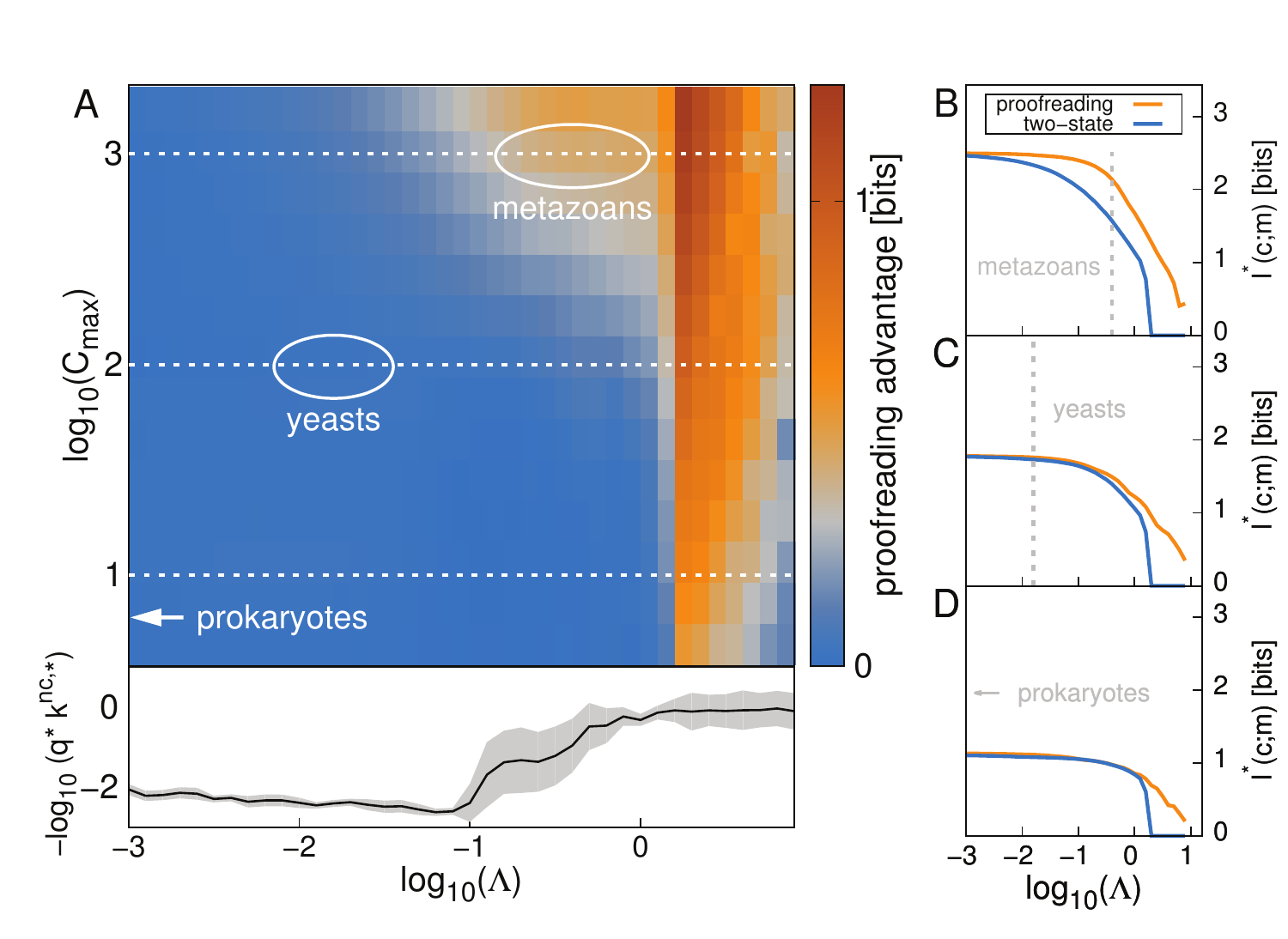}
\caption{{\bf A)} Information advantage (in bits, color scale) of optimal proofreading over optimal two-state architectures, as a function of crosstalk severity $\Lambda$ and dynamic range of  input TF concentration, $C_{\rm max}$. Typical values for prokaryotes, yeast, and metazoans are marked in white. Lower inset: optimal rates, $q^*k_-^{\rm nc,*}$ (black line = average over $C_{\rm max}$, gray shade = std), indicate a switch to the proofreading strategy. {\bf B, C, D)} Cuts through the information plane in (A) along white dashed lines showing the collapse of two-state performance as $\log_{10}(\Lambda)\rightarrow 0$ and a clear proofreading advantage for metazoan regulation.}
\label{f3}
\end{figure}

Figure~\ref{f3}A shows the advantage, in bits, of the optimal proofreading architecture relative to the optimal non-proofreading one. This ``information plane,'' $I_{q*}(m;c)-I_{q=0}(m;c)$, is plotted as a function of $\Lambda$ and $C_{\rm max}$. In the limit $\Lambda\rightarrow 0$, the difference in performance goes to zero: there, optimization drives $q^*k_-^{\rm nc,*}\gg 1$, but proofreading offers vanishing advantage over the optimal two-state promoter architecture when noncognate binding is negligible. 
As $\Lambda$ increases, proofreading becomes beneficial over the two-state architecture, and more so for higher values of $C_{\rm max}$. Higher input concentrations $c\in [0,C_{\rm max}]$ permit faster on-rates, resulting in faster optimal off rates $k_-^{\rm c,*}$ and faster optimal $1/q^*$. Generally, faster switching of promoter states in Fig~\ref{f1}B means that promoter switching noise will be lower and thus information higher (at fixed mean mRNA expression $\bar{m}$); in particular, optimization tends to minimize promoter switching noise by selecting the fastest $1/q$ that still admits error rejection, i.e., $q^*k_-^{\rm nc,*}\sim 1$. At $\Lambda=\nu/\sigma\simeq 1$, the signaling capacity of the non-proofreading architecture collapses completely, with $I_{q=0}(c;m)\approx 0$ \footnote{This is independent of whether one modulates $\Lambda$ by changing $\nu$, as for Fig~\ref{f3}A, or by changing $\sigma$; although the optimal rates may take on different values, the information plane is essentially unchanged irrespective of how $\Lambda$ is modulated.}. At this point optimal proofreading architectures are affected, but still generally maintain at least half of the capacity seen at $\Lambda=0$; proofreading extends the performance of the gene regulation well into the $\Lambda>0$ region, before finally succumbing to crosstalk. 

Where do different organisms lie in the information plane? Prokaryotes have on the order of $\nu\sim 100$ types of transcription factors, whose binding site motifs typically contain  around $23$ bits of sequence information \cite{wunderlich+mirny_09}, corresponding to the binding energy difference of $16\,k_BT$ between cognate and noncognate sites \cite{gerland+hwa}, and thus a specificity of roughly $\sigma\sim10^7$. This corresponds to a small value of  crosstalk severity, $\Lambda\sim 10^{-5}$. For yeast, the typical sequence information is $14$ bits ($10\,k_BT$) \cite{wunderlich+mirny_09}, which gives $\Lambda\sim 0.01$ (for $\nu \sim 200$ \cite{babu}). For multicellular eukaryotes, the typical sequence information is $12$ bits ($8\,k_B T$), and the number of TF species varies between $\nu\approx 10^3$ (\emph{C. elegans}) to $\nu \approx 2\cdot 10^3$  (human) \cite{bionumbers}, putting $\Lambda $ between 0.1 and 1. We can also estimate the dimensionless parameter $C_{\rm max}=k_+ C /d$. Assuming diffusion-limited binding of TFs to their binding sites, $k_+ C / d \approx 3 D a N / R^3 d$, where $D\sim 1\mathrm{\mu m^3/s}$ is the typical TF diffusion constant \cite{bionumbers}, $a\sim 3\e{nm}$ is the binding site size, $R=3\e{\mu m}$ ($1\e{\mu m}$) is the radius of an eukaryotic nucleus (prokaryotic cell), and $N$ is the typical copy number of TFs per nucleus ($N\sim 10$ for prokaryotes, $10^3$ for yeast, $10^3-10^5$ for eukaryotes). Typical mRNA lifetimes are $5-10$ min in prokaryotes, $20-30$ min in yeast, and $>1$ hour in metazoans. This yields $C_{\rm max}$ of order 10 for prokaryotes, $10^2$ for yeast cells, and $>10^3$ for multicellular eukaryote cells. While these are very rough estimates, different kinds of cells clearly differ substantially in their location on the information plane of Fig~\ref{f3}A.

Taken together, these values suggest that crosstalk is acute for metazoans and that proofreading in gene regulation could provide a vast improvement over equilibrium regulation schemes, as in Fig~\ref{f3}B. In contrast, our proposal offers no advantage for prokaryotes, and remains agnostic about yeast (Figs~\ref{f3}C, D). While much remains unknown about the molecular machinery of eukaryotic gene regulation, it has been experimentally shown that transcriptional initiation (not just elongation) involves a series of out-of-equilibrium steps. Amongst those, perhaps the most intriguing are the covalent modifications on the eukaryotic RNA polymerase II CTD tail \cite{egloff+murphy}. The tail contains tandem repeats of short peptides (from 26 repeats in yeast to 52 in mammals), which need to get phosphorylated in order to initiate transcription and subsequently cleared after completed transcription in order to reuse the polymerase; genetic interference with this tail seems to be lethal. One can contemplate a scenario where a sequence of such phosphorylation steps corresponds to the out-of-equilibrium reaction $q$ of our simple proofreading scheme, ``ticking away'' time until the polymerase commits to initiation, with every tick giving the machinery another opportunity to check if cognate TFs are still bound and, if not, abort transcription. The existence of any such (or similar) proofreading scheme would be interesting, but is currently purely hypothetical.

Why would eukaryotes employ a method of gene regulation so qualitatively different from prokaryotes, instead of simply using longer, specific binding sites that would drive crosstalk severity $\Lambda$ towards zero? While beyond the scope of this work, one possible hypothesis is that such longer sites are not easily evolvable and, additionally, that the complexity of regulation calls for combinatorial control of single genes by many TFs of different species, each of which could have weak specificity. Such cooperative or combinatorial control could indeed address a specific target gene uniquely, as proposed (e.g., \cite{todeschini+al_14,wunderlich+mirny_09}); what has largely been neglected in previous discussions is that it would be difficult to \emph{prevent} the target gene from being erroneously induced by crosstalk. Here we advanced a possible hypothetical mechanism, proofreading-based transcriptional regulation, to mitigate this problem. It is interesting to note that, unlike most biophysical problems where we clearly appreciate  their out-of-equilibrium nature, transcriptional regulation has remained a textbook example of a non-trivial \emph{equilibrium} molecular recognition process, likely due to the success of the equilibrium assumption in prokaryotes. Perhaps constraints imposed by crosstalk will motivate us to reexamine this assumption in eukaryotic regulation more closely.

\begin{acknowledgments}
We thank TR Sokolowski and T Friedlander for helpful comments on the manuscript, and E van Nimwegen for suggesting that histone modification / remodeling might also constitute a candidate proofreading mechanism.
\end{acknowledgments}
%
%

%
 \end{document}